\documentclass[english]{IEEEtran}
\usepackage[T1]{fontenc}
\usepackage[latin9]{inputenc}
\usepackage{color}
\usepackage{array}
\usepackage{float}
\usepackage{booktabs}
\usepackage{multirow}
\usepackage{amsmath}
\usepackage{amsthm}
\usepackage{amssymb}
\usepackage{graphicx}

\makeatletter

\providecommand{\tabularnewline}{\\}
\floatstyle{ruled}
\newfloat{algorithm}{tbp}{loa}
\providecommand{\algorithmname}{Algorithm}
\floatname{algorithm}{\protect\algorithmname}

\theoremstyle{plain}
\newtheorem{thm}{\protect\theoremname}

\usepackage{balance}
\usepackage{babel}
\usepackage{multirow}
\usepackage{amsmath}
\usepackage{graphicx}
\usepackage{bigstrut}
\renewcommand{\fnum@figure}{Fig.~\thefigure} 
\usepackage{cite}
\usepackage{adjustbox}
\usepackage{color}
\usepackage{soul}
\usepackage{url}
\usepackage{stackrel}

\makeatother

\usepackage{babel}
\usepackage{listings}
\providecommand{\theoremname}{Theorem}

\begin{document}

\title{Real-Time Equality-Constrained Hybrid State Estimation in Complex
Variables}

\author{Izudin Džafi\'{c}, \emph{Senior Member}, \emph{IEEE}, Rabih A. Jabr,
\emph{Fellow}, \emph{IEEE}, and Bikash C. Pal,\emph{ Fellow}, \emph{IEEE}
\thanks{I. Džafi\'{c} is with the International University of Sarajevo, Hrasni\v{c}ka
cesta 15, 71210 Sarajevo, Bosnia (email: idzafic@ieee.org).}\thanks{R. A. Jabr is with the Department of Electrical \& Computer Engineering,
American University of Beirut, P.O. Box 11-0236, Riad El- Solh / Beirut
1107 2020, Lebanon (email: rabih.jabr@aub.edu.lb).}\thanks{B. C. Pal is with the Electrical and Electronic Engineering Department,
Imperial College, London SW7 2AZ, U.K. (e-mail: b.pal@imperial.ac.uk).}}
\maketitle
\begin{abstract}
The hybrid power system state estimation problem requires computing
the state of the power network using data from both legacy and phasor
measurements. Recent research has shown that the normal equations
approach in complex variables is computationally advantageous, particularly
in the presence of phasor measurement values, and that its software
implementation is best suited to modern processors that employ single
instruction multiple data (SIMD) processor extensions. The complex
normal equations approach is however not ideal for handling zero injection
measurements, as it requires their modeling as pseudo-measurements
with high weights. This paper employs Wirtinger calculus for extending
the complex normal equations approach to include equality constraints,
and contrasts it with two previously published implementations: the
normal equations approach in complex variables and the hybrid equality
constrained state estimator in real variables. Numerical results are
reported on transmission networks having up to 9241 nodes; they show
that the complex variable equality constrained hybrid state estimator
exhibits superior performance as compared to the above two techniques
in terms of both computational time and accuracy. Moreover, the execution
time on the largest network is less than 300 ms, which makes the proposed
implementation commensurate with the requirements of real-time applications.
\end{abstract}


\section{Introduction}

\IEEEPARstart{T}{he }power system state estimation problem is most
commonly formulated as a weighted least squares problem and solved
via the normal equations (NE) approach; the main advantage of the
NE approach is that it gives rise to a gain matrix that can be rapidly
factorized using sparsity techniques. Nodes that have neither generation
nor load are modeled as virtual zero complex power injection measurements,
which are very useful to enhance the estimation accuracy. Zero injection
measurements can be accounted for in the NE approach by assigning
relatively much larger weights to them, but this can potentially cause
ill-conditioning. It is now accepted that zero injections are best
handled through formulating the estimation problem as an equality
constrained optimization program, leading to methods such as the normal
equations with equality constraints, Hatchel's augmented matrix method,
and other hybrid forms and extensions \cite{Nucera_1991,Abur_B_2004,Gomez_Quiles_2013}.

In the current operational practice, state estimators are expected
to employ measurements from both the supervisory control and data
acquisition (SCADA) and the phasor measurement unit (PMU) systems.
PMU measurements are increasingly being employed in power systems,
with benefits in risk mitigation against cyber attacks \cite{Taha_2018},
in Volt-VAr control \cite{Borghetti_2017}, and in transmission network
state estimation \cite{Xu_2017}. Handling phasor current measurements
in power grid state estimation can be accomplished using either noninvasive
or direct techniques. The noninvasive methods have the distinct advantage
of not requiring any update to the classical state estimator implementations
that are based on SCADA measurements; they rather employ post processing
techniques that improve the estimation outcome by fusing the SCADA
measurement-based state vector with the PMU measurements \cite{Zhou_2006,Nuqui_2007,Cheng_2008,Baltensperger_2010,SimoesCosta_2013,Wu_2017}.
Noninvasive methods however require complete observability by the
SCADA telemetering system. On the other hand, the direct techniques
do not necessitate SCADA system observability, but rather treat both
SCADA and PMU measurements in a unified optimization framework \cite{Bi_2008,Chakrabarti_2010,Valverde_2011}.
The large disparity in data refresh rates amongst PMU and SCADA measurements
is an issue common to all hybrid state estimators; the proposed solutions
include buffering PMU measurements \cite{Murugesan_2015} and SCADA
state reconstruction techniques \cite{Glavic_2013,Gol_2105H}. 

Recent research has shown that hybrid state estimation can be carried
out using the NE approach in complex variables \cite{Dzafic_2018},
and that the complex variable approach has advantages (i) in handling
complex-valued measurements and (ii) in implementation on modern processors
that support single instruction multiple data (SIMD) operations \cite{SIMD}.
SIMD instructions allow processing multiple pieces of data using a
single instruction, thus speeding up the throughput of implementations
for video encoding/decoding, image processing, and data analysis \cite{AVX2};
SIMD instruction sets also allow fused multiply-accumulate operations,
which could naturally be leveraged in complex variable computing applications.
The complex NE (CNE) approach in \cite{Dzafic_2018} is a generalization
of the real variable implementation, derived via Wirtinger calculus
\cite{wirtinger,Delgado_2009}. This paper extends the CNE approach
to include equality constraints, and thus effectively handle zero
injection measurements for further improving the estimation accuracy;
the complex equality constrained (CEC) estimator is implemented using
advanced vector extensions (AVX-2) \cite{AVX2} and contrasted with
both the CNE approach \cite{Dzafic_2018} and the real equality constrained
(REC) hybrid state estimator \cite{Valverde_2011}. 

The rest of this paper is organized as follows. Section \ref{sec:Complex-Normal-Equations}
reviews the normal equations approach in complex variables, and Section
\ref{sec:Complex-Equality-Constrained} presents the extension to
the complex variable normal equations with equality constraints. An
introduction to the program implementation via AVX-2 is given in Section
\ref{sec:AVX-2}. Section \ref{sec:Numerical-Results-} presents numerical
results and comparisons with the CNE \cite{Dzafic_2018} and REC \cite{Valverde_2011}
estimators on networks with up to 9241 nodes. The paper is concluded
in Section \ref{sec:Conclusion}.

\section{Complex Normal Equations (CNE) \label{sec:Complex-Normal-Equations}}

Consider the power system state estimation problem in complex variables
(\ref{eq:inverse_problem}), where {\small{}$h(x,\overline{x})$}
is the vector of measurement functions, {\small{}$z$} is the vector
of measured values, {\small{}$x$} is the vector of complex state
variables (phasor voltages), and {\small{}$\overline{x}$} denotes
the conjugate of {\small{}$x$} \cite{Dzafic_2018}:

{\small{}
\begin{equation}
h(x,\overline{x})\thickapprox z\label{eq:inverse_problem}
\end{equation}
}{\small \par}

{\small{}
\begin{equation}
h(x,\overline{x})=\left[\begin{array}{c}
h_{1}(x,\overline{x})\\
\vdots\\
h_{m}(x,\overline{x})
\end{array}\right],\quad z=\left[\begin{array}{c}
z_{1}\\
\vdots\\
z_{m}
\end{array}\right]
\end{equation}
}Using Wirtinger calculus, (\ref{eq:inverse_problem}) can be linearized
via the complex Taylor series expansion around the current estimate
of the state vector {\small{}$\left[x^{*};\overline{x}^{*}\right]$};
{\small{}$H$} is formed by the Jacobian {\small{}$H_{x}$} and the
conjugate Jacobian {\small{}$H_{\overline{x}}$} matrices evaluated
at {\small{}$\left[x^{*};\overline{x}^{*}\right]$}:

{\small{}
\begin{equation}
h(x^{\star},\overline{x}^{\star})+H\left[\begin{array}{c}
\Delta x\\
\Delta\overline{x}
\end{array}\right]\approx z
\end{equation}
}{\small \par}

{\small{}
\begin{equation}
H=\left[H_{x},H_{\overline{x}}\right]=\left[\begin{array}{cccccc}
\frac{\partial h_{1}}{\partial x_{1}} & \cdots & \frac{\partial h_{1}}{\partial x_{n}} & \frac{\partial h_{1}}{\partial\overline{x}_{1}} & \cdots & \frac{\partial h_{1}}{\partial\overline{x}_{n}}\\
\vdots & \ddots & \vdots & \vdots & \ddots & \vdots\\
\frac{\partial h_{m}}{\partial x_{1}} & \cdots & \frac{\partial h_{m}}{\partial x_{n}} & \frac{\partial h_{m}}{\partial\overline{x}_{1}} & \cdots & \frac{\partial h_{m}}{\partial\overline{x}_{n}}
\end{array}\right]
\end{equation}
}The correction to the state vector {\small{}$\left[\Delta x^{*};\Delta\overline{x}^{*}\right]$}
is obtained by minimizing the weighted least squares (WLS) objective
value: 

{\small{}
\begin{equation}
\ell(\Delta x,\Delta\overline{x})=\frac{1}{2}\left(\overline{r}-\overline{H}\left[\begin{array}{c}
\Delta\overline{x}\\
\Delta x
\end{array}\right]\right)^{T}W\left(r-H\left[\begin{array}{c}
\Delta x\\
\Delta\overline{x}
\end{array}\right]\right)\label{eq:obj_LS_0}
\end{equation}
}where {\small{}$W$} is a diagonal matrix of measurement weights
and {\small{}$r=z-h(x^{\star},\overline{x}^{\star})$}. Eq. (\ref{eq:obj_LS_0})
can be expanded into (\ref{eq:obj_LS_1}), with the vector {\small{}$\beta$}
and matrix {\small{}$G$} given by (\ref{eq:beta}) and (\ref{eq:G_matrix}),
respectively:

{\small{}
\begin{equation}
\begin{aligned}\ell & =\frac{1}{2}\left(\overline{r}^{T}Wr-\overline{\beta}^{T}\left[\begin{array}{c}
\Delta x\\
\Delta\overline{x}
\end{array}\right]-\left[\begin{array}{cc}
\Delta\overline{x}^{T} & \Delta x^{T}\end{array}\right]\beta\right)\\
 & +\frac{1}{2}\left(\left[\begin{array}{cc}
\Delta\overline{x}^{T} & \Delta x^{T}\end{array}\right]G\left[\begin{array}{c}
\Delta x\\
\Delta\overline{x}
\end{array}\right]\right)
\end{aligned}
\label{eq:obj_LS_1}
\end{equation}
}{\small \par}

{\small{}
\begin{equation}
\beta=\overline{H}^{T}Wr=\left[\begin{array}{c}
\beta_{\overline{x}}\\
\beta_{x}
\end{array}\right]\label{eq:beta}
\end{equation}
\begin{equation}
G=\overline{H}^{T}WH=\left[\begin{array}{cc}
G_{\overline{x}x} & G_{\overline{xx}}\\
G_{xx} & G_{x\overline{x}}
\end{array}\right]\label{eq:G_matrix}
\end{equation}
}Ref. \cite{Dzafic_2018} shows that for hybrid power system state
estimation, the elements of {\small{}$G$} and {\small{}$\beta$}
satisfy the following properties: {\small{}$\beta_{x}=\overline{\beta}_{\overline{x}}$},
{\small{}$G_{x\overline{x}}=\overline{G}_{\overline{x}x}=G_{\overline{x}x}^{T}$},
and {\small{}$G_{xx}=G_{xx}^{T}=\overline{G}_{\overline{xx}}=\overline{G}_{\overline{xx}}^{T}$}.
Then the minimizer of (\ref{eq:obj_LS_1}) is given by the solution
of the normal equations:

{\small{}
\begin{equation}
G\left[\begin{array}{c}
\Delta x\\
\Delta\overline{x}
\end{array}\right]=\beta\label{eq:Theorem_res}
\end{equation}
}{\small \par}

\section{Complex Equality Constrained (CEC) Normal Equations \label{sec:Complex-Equality-Constrained}}

\begin{figure}[tb]
\noindent \begin{centering}
\rotatebox{-90}{\includegraphics[height=6.5cm]{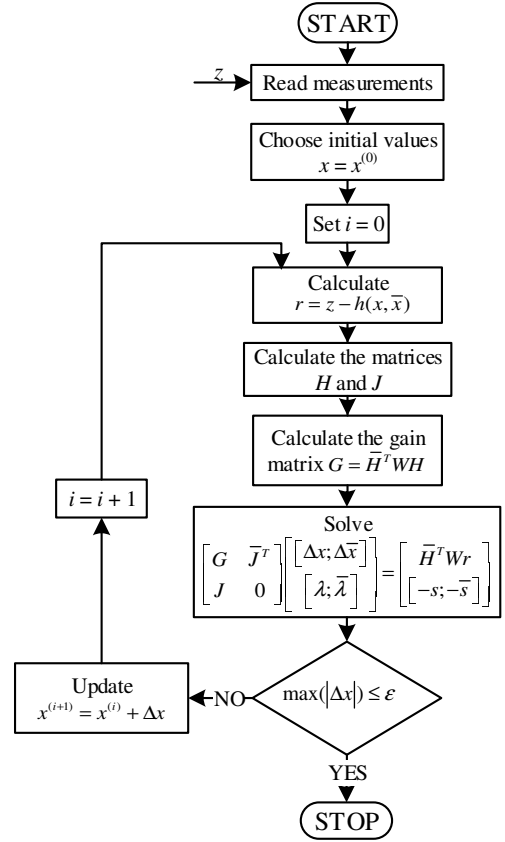}}
\par\end{centering}
\caption{Hybrid SE via Complex Equality Constrained (CEC) Normal Equations.\label{fig:CEC}}
\end{figure}

The zero injection measurements give rise to a large disparity of
weights in the normal equations approach, and may lead to severe ill-conditioning
\cite{Nucera_1991}. This problem can be alleviated by a constrained
WLS method, which treats zero injection measurements as equality constraints.
The zero injection at node {\small{}$i$} is modeled by the complex
power injection ({\small{}$s_{i}(x,\overline{x})=0$)} and its conjugate
({\small{}$\overline{s}_{i}(x,\overline{x})=0$}), leading to the
following relationship between their Wirtinger derivatives \cite{Delgado_2009}:

{\small{}
\begin{equation}
\frac{\partial\overline{s}_{i}}{\partial x_{j}}=\overline{\left(\frac{\partial s_{i}}{\partial\overline{x}_{j}}\right)},\quad\frac{\partial\overline{s}_{i}}{\partial\overline{x}_{j}}=\overline{\left(\frac{\partial s_{i}}{\partial x_{j}}\right)}
\end{equation}
}Therefore, the constrained linear WLS problem can be written as:
{\small{}
\begin{equation}
\min\ell(\Delta x,\Delta\overline{x})\label{eq:obj_LS_2}
\end{equation}
}subject to:{\small{}
\begin{equation}
J\left[\begin{array}{c}
\Delta x\\
\Delta\overline{x}
\end{array}\right]=\left[\begin{array}{cc}
J_{x} & J_{\overline{x}}\\
\overline{J}_{\overline{x}} & \overline{J}_{x}
\end{array}\right]\left[\begin{array}{c}
\Delta x\\
\Delta\overline{x}
\end{array}\right]=\left[\begin{array}{c}
-s\\
-\overline{s}
\end{array}\right]\label{eq:constr_LS2}
\end{equation}
}where the elements in the complex vectors {\small{}$s$} and {\small{}$\overline{s}$}
contain the equations for the zero complex power injections and their
conjugates, except for the last element in each vector that sets the
slack angle condition; the corresponding equations are given in the
Appendix. The classical theory of Lagrange multipliers for solving
constrained minimization problems stipulates that the objective function
and constraints are real-valued functions of real unknown variables;
however by applying Wirtinger calculus, \cite{Delgado_2009} shows
that a stationary point of the Lagrangian function (\ref{eq:Lagrangian})
is a solution to (\ref{eq:obj_LS_2})-(\ref{eq:constr_LS2}), where
{\small{}$\Re$} denotes the real part of a complex quantity:

{\small{}
\begin{equation}
\mathcal{L=}\ell(x,\overline{x})+\Re\left\{ \left[\overline{\lambda}^{T},\lambda^{T}\right]\left(\left[\begin{array}{cc}
J_{x} & J_{\overline{x}}\\
\overline{J}_{\overline{x}} & \overline{J}_{x}
\end{array}\right]\left[\begin{array}{c}
\Delta x\\
\Delta\overline{x}
\end{array}\right]+\left[\begin{array}{c}
s\\
\overline{s}
\end{array}\right]\right)\right\} \label{eq:Lagrangian}
\end{equation}
}{\small \par}
\begin{thm}
The solution to the complex constrained linear WLS problem (\ref{eq:obj_LS_2})-(\ref{eq:constr_LS2})
that arises in hybrid power system state estimation is given by the
normal equations with equality constraints:

{\small{}
\begin{equation}
\left[\begin{array}{cccc}
G_{\overline{x}x} & G_{\overline{xx}} & \overline{J}_{x}^{T} & J_{\overline{x}}^{T}\\
\overline{G}_{\overline{xx}} & \overline{G}_{\overline{x}x} & \overline{J}_{\overline{x}}^{T} & J_{x}^{T}\\
J_{x} & J_{\overline{x}} & 0 & 0\\
\overline{J}_{\overline{x}} & \overline{J}_{x} & 0 & 0
\end{array}\right]\left[\begin{array}{c}
\Delta x\\
\Delta\overline{x}\\
\lambda\\
\overline{\lambda}
\end{array}\right]=\left[\begin{array}{c}
\beta_{\overline{x}}\\
\overline{\beta}_{\overline{x}}\\
-s\\
-\overline{s}
\end{array}\right]\label{eq:res_theorem}
\end{equation}
}{\small \par}
\end{thm}
\begin{IEEEproof}
Eq. (\ref{eq:real_part}) shows that the function in parenthesis in
(\ref{eq:Lagrangian}) is real:

{\small{}
\begin{align}
\left[\overline{\lambda}^{T},\lambda^{T}\right]\left(\left[\begin{array}{cc}
J_{x} & J_{\overline{x}}\\
\overline{J}_{\overline{x}} & \overline{J}_{x}
\end{array}\right]\left[\begin{array}{c}
\Delta x\\
\Delta\overline{x}
\end{array}\right]+\left[\begin{array}{c}
s\\
\overline{s}
\end{array}\right]\right) & =\nonumber \\
\underset{2\Re\left\{ \overline{\lambda}^{T}J_{x}\Delta x\right\} }{\underbrace{\overline{\lambda}^{T}J_{x}\Delta x+\lambda^{T}\overline{J}_{x}\Delta\overline{x}}}+\underset{2\Re\left\{ \overline{\lambda}^{T}J_{\overline{x}}\Delta\overline{x}\right\} }{\underbrace{\overline{\lambda}^{T}J_{\overline{x}}\Delta\overline{x}+\lambda^{T}\overline{J}_{\overline{x}}\Delta x}}+\underset{2\Re\left\{ \overline{\lambda}^{T}s\right\} }{\underbrace{\overline{\lambda}^{T}s+\lambda^{T}\overline{s}}}\label{eq:real_part}
\end{align}
}Using (\ref{eq:real_part}), the Lagrangian function (\ref{eq:Lagrangian})
reduces to:{\small{}
\begin{equation}
\mathcal{L=}\ell(x,\overline{x})+\left[\overline{\lambda}^{T},\lambda^{T}\right]\left(\left[\begin{array}{cc}
J_{x} & J_{\overline{x}}\\
\overline{J}_{\overline{x}} & \overline{J}_{x}
\end{array}\right]\left[\begin{array}{c}
\Delta x\\
\Delta\overline{x}
\end{array}\right]+\left[\begin{array}{c}
s\\
\overline{s}
\end{array}\right]\right)
\end{equation}
}Therefore:{\small{}
\begin{align}
\nabla_{\Delta\overline{x}}\mathcal{L} & =\nabla_{\Delta\overline{x}}\ell+\left[\overline{\lambda}^{T}J_{\overline{x}}+\lambda^{T}\overline{J}_{x}\right]^{T}\nonumber \\
 & =-\beta_{\overline{x}}+G_{\overline{x}x}\Delta x+G_{\overline{xx}}\Delta\overline{x}+\overline{J}_{x}^{T}\lambda+J_{\overline{x}}^{T}\overline{\lambda}\label{eq:grad_Dx_bar}
\end{align}
\begin{align}
\nabla_{\Delta x}\mathcal{L} & =\nabla_{\Delta x}\ell+\left[\overline{\lambda}^{T}J_{x}+\lambda^{T}\overline{J}_{\overline{x}}\right]^{T}\nonumber \\
 & =-\overline{\beta}_{\overline{x}}+\overline{G}_{\overline{xx}}\Delta x+\overline{G}_{\overline{x}x}\Delta\overline{x}+\overline{J}_{\overline{x}}^{T}\lambda+J_{x}^{T}\overline{\lambda}\label{eq:grad_Dx}
\end{align}
}Equating (\ref{eq:grad_Dx_bar}) and (\ref{eq:grad_Dx}) to zero,
together with the feasibility constraints (\ref{eq:constr_LS2}),
results in a set of equations that is necessary and sufficient to
compute a stationary point of {\small{}$\mathcal{L}$} (a minimizer
of (\ref{eq:obj_LS_2})-(\ref{eq:constr_LS2}) due to the convexity
of the problem); these conditions are given in (\ref{eq:res_theorem}).
It stays to demonstrate that the solution to (\ref{eq:res_theorem})
gives two pairs of complex conjugate vectors, {\small{}$\left[\Delta x;\Delta\overline{x}\right]$}
and {\small{}$\left[\lambda;\overline{\lambda}\right]$}. To show
this, write (\ref{eq:res_theorem}) as:

{\small{}
\begin{equation}
\left[\begin{array}{cccc}
G_{\overline{x}x} & G_{\overline{xx}} & \overline{J}_{x}^{T} & J_{\overline{x}}^{T}\\
\overline{G}_{\overline{xx}} & \overline{G}_{\overline{x}x} & \overline{J}_{\overline{x}}^{T} & J_{x}^{T}\\
J_{x} & J_{\overline{x}} & 0 & 0\\
\overline{J}_{\overline{x}} & \overline{J}_{x} & 0 & 0
\end{array}\right]\left[\begin{array}{c}
\Delta x\\
\Delta y\\
\lambda\\
\mu
\end{array}\right]=\left[\begin{array}{c}
\beta_{\overline{x}}\\
\overline{\beta}_{\overline{x}}\\
-s\\
-\overline{s}
\end{array}\right]\label{eq:res_theorem-2}
\end{equation}
}Taking the complex conjugate of (\ref{eq:res_theorem-2}) gives:{\small{}
\begin{equation}
\left[\begin{array}{cccc}
\overline{G}_{\overline{x}x} & \overline{G}_{\overline{xx}} & J_{x}^{T} & \overline{J}_{\overline{x}}^{T}\\
G_{\overline{xx}} & G_{\overline{x}x} & J_{\overline{x}}^{T} & \overline{J}_{x}^{T}\\
\overline{J}_{x} & \overline{J}_{\overline{x}} & 0 & 0\\
J_{\overline{x}} & J_{x} & 0 & 0
\end{array}\right]\left[\begin{array}{c}
\Delta\overline{x}\\
\Delta\overline{y}\\
\overline{\lambda}\\
\overline{\mu}
\end{array}\right]=\left[\begin{array}{c}
\overline{\beta}_{\overline{x}}\\
\beta_{\overline{x}}\\
-\overline{s}\\
-s
\end{array}\right]\label{eq:res_theorem-3}
\end{equation}
}Swapping as a whole the first row with the second (in (\ref{eq:res_theorem-3})),
the third row with the fourth, the first column with the second, and
the third column with the fourth gives:

{\small{}
\begin{equation}
\left[\begin{array}{cccc}
G_{\overline{x}x} & G_{\overline{xx}} & \overline{J}_{x}^{T} & J_{\overline{x}}^{T}\\
\overline{G}_{\overline{xx}} & \overline{G}_{\overline{x}x} & \overline{J}_{\overline{x}}^{T} & J_{x}^{T}\\
J_{x} & J_{\overline{x}} & 0 & 0\\
\overline{J}_{\overline{x}} & \overline{J}_{x} & 0 & 0
\end{array}\right]\left[\begin{array}{c}
\Delta\overline{y}\\
\Delta\overline{x}\\
\overline{\mu}\\
\overline{\lambda}
\end{array}\right]=\left[\begin{array}{c}
\beta_{\overline{x}}\\
\overline{\beta}_{\overline{x}}\\
-s\\
-\overline{s}
\end{array}\right]\label{eq:res_theorem-1}
\end{equation}
}Now comparing (\ref{eq:res_theorem-1}) with (\ref{eq:res_theorem-2})
shows that {\small{}$\Delta y=\Delta\overline{x}$} and {\small{}$\mu=\overline{\lambda}$,}
i.e. the solution to (\ref{eq:res_theorem}) gives a pair of complex
conjugate solutions and is therefore admissible.
\end{IEEEproof}
Theorem 1 reveals that the complex normal equations with equality
constraints (CEC) has a form analogous to the real variable EC estimator;
the flowchart for the CEC method is given in Fig. \ref{fig:CEC}.
The Appendix of this paper shows the elements of the {\small{}$J$}
matrix, whereas the elements of {\small{}$H$} that are required in
forming the {\small{}$G$ matrix} are available in \cite{Dzafic_2018}.

\section{Advanced Vector Extensions: AVX-2 \label{sec:AVX-2}}

Modern processors support single instruction multiple data (SIMD)
processor extensions, which include Advanced Vector Extensions - AVX2
\cite{AVX2}. AVX2 uses 256-bit registers and therefore allows the
manipulation of two double precision complex values (two real and
two imaginary parts) per register of CPU; this results in fast arithmetic
operations with complex numbers, and makes AVX-2 ideal for implementing
real-time estimation and control functions in complex variables.

\begin{figure}
\centering{}\includegraphics[width=7.8cm]{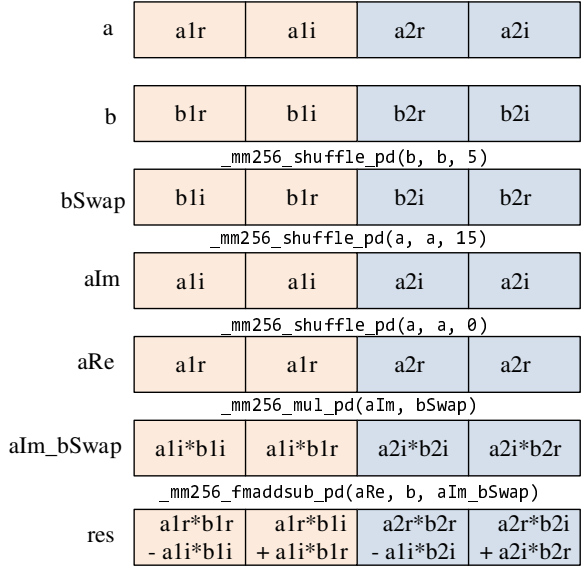}\caption{Vectorized complex multiplication using AVX-2.\label{fig:Vectorized-complex-multiplicatio}}
\end{figure}

Consider for illustration the product of two complex numbers:{\small{}
\begin{align}
\mathrm{\mathrm{\left(a1r+\mathtt{i}*a1i\right)}*\left(\mathrm{b1r+\mathtt{i}*b1i}\right)}=\nonumber \\
\mathrm{\left(a1r*b1r-a1i*b1i\right)+\mathtt{i}*\left(a1r*b1i+a1i*b1r\right)}\label{eq:cmplx-product}
\end{align}
}From an implementation perspective, the non-vectorized multiplication
in (\ref{eq:cmplx-product}) would require 4 multiplication steps,
1 addition step, and 1 subtraction step; when performing two complex
number multiplications {\small{}$\left(\mathrm{a1r+\mathtt{i}*a1i}\right)*\left(\mathrm{b1r+\mathtt{i}*b1i}\right)$}
and {\small{}$\left(\mathrm{a2r+\mathtt{i}*a12}\mathrm{i}\right)*\left(\mathrm{b2r+\mathtt{i}*b2i}\right)$},
the non-vectorized code would therefore require 8 multiplication steps,
2 addition steps, and 2 subtraction steps. In contrast, the AVX-2
code in Algorithm \ref{alg:Computation-of-two} performs the same
multiplication but using 2 multiplication steps (each step involves
4 simultaneous multiplications), 1 additions/subtraction step, and
3 register shuffles. Fig. \ref{fig:Vectorized-complex-multiplicatio}
illustrates the corresponding steps in the 256-bit CPU registers.
In Fig. \ref{fig:Vectorized-complex-multiplicatio}, {\small{}$\mathrm{a}$}
is a 256-bit register holding two complex numbers ({\small{}$\mathrm{\left(a1r+\mathtt{i}*a1i\right)}$}
and {\small{}$\left(\mathrm{a2r+\mathtt{i}*a2i}\right)$}), where
each of the real and imaginary components are stored in a 64-bit register
block; similarly, the 256-bit register {\small{}$\mathrm{b}$} contains
two complex numbers ({\small{}$\left(\mathrm{b1r+\mathtt{i}*b1i}\right)$}
and {\small{}$\left(\mathrm{b2r+\mathtt{i}*b2i}\right)$}). To achieve
the multiplication, four intermediate results are stored in four 256-bit
registers: register {\small{}$\mathrm{bSwap}$} contains the swapped
elements of {\small{}$\mathrm{b}$}, registers {\small{}$\mathrm{aIm}$}
and {\small{}$\mathrm{aRe}$} contain only duplicates of the imaginary
and real parts of {\small{}$\mathrm{a}$}, and the register {\small{}$\mathrm{aIm}$\_$\mathrm{\mathrm{b}Swap}$}
holds the block register multiplication of {\small{}$\mathrm{aIm}$}
and {\small{}$\mathrm{b}\mathrm{Swap}$}. The result of the multiplication
is in the register {\small{}$\mathrm{res}$}; it is formed using the
fused multiply add/subtract function which takes the multiplication
of {\small{}$\mathrm{aRe}$} and {\small{}$\mathrm{\mathrm{b}}$},
and adds/subtracts the block registers (2 and 4)/(1 and 3) of {\small{}$\mathrm{aIm}$\_$\mathrm{\mathrm{b}Swap}$}.
The fused multiply add/subtract function is particularly targeted
to complex number operations; it allows faster multiplication of two
complex (double precision) numbers by two complex numbers.

AVX-2 includes several advances related to the new fused-multiply-add
(FMA) instructions; these have been leveraged in the implementation
of the CEC estimator in Fig. \ref{fig:CEC}.

\begin{algorithm}
\begin{lstlisting}[language={C++},basicstyle={\footnotesize},tabsize=4]
// AVX2 with fused-multiply-add intrinsics 
// requires only two multiplications to multiply 
// two double precision complex numbers

__m256d mult(__m256d const& a, __m256d const& b)
{         
	// Swap b.re and b.im 
	__m256d bSwap = _mm256_shuffle_pd(b,b,5);      
	// Imag part of a in both   
	__m256d aIm = _mm256_shuffle_pd(a,a,15);
	// Real part of a in both        
	__m256d aRe = _mm256_shuffle_pd(a,a,0);  
	//First multiplication (a.im*b.im, a.im*b.re) 
	__m256d aIm_bSwap = _mm256_mul_pd(aIm, bSwap); 
	//Second multiplication with fused add/sub
	// aRe*b + complex(-aIm_bSwap.re,aIm_bSwap.im) 
	return  _mm256_fmaddsub_pd(aRe, b, aIm_bSwap);
}
\end{lstlisting}

\caption{Computation of two complex values using double precision AVX-2 with
Fuse-Multiply-Add intrinsics\label{alg:Computation-of-two}}
\end{algorithm}

\section{Numerical Results \label{sec:Numerical-Results-} }

The CEC state estimator was programmed in C++, and the computations
were performed using solvers developed via the AVX-2 processor extension
\cite{AVX2}; comparative analysis was carried out with the CNE implementation
in \cite{Dzafic_2018} (see section II), and with an implementation
of the real variable equality constrained (REC) hybrid state estimator
for handling both SCADA and PMU measurements \cite{Valverde_2011}.
The numerical tests were executed on a PC with an Intel i5-7600K processor
and 16 GB of RAM. The termination tolerance {\small{}$\varepsilon$}
in Fig. \ref{fig:CEC} was set to {\small{}$10^{-6}$} per-unit. The
testing was carried out on the IEEE 118 test network (118), the French
very high-voltage 1888 node network (1888), and part of the of the
European high voltage transmission network with 9241 nodes \cite{Josz_2016}.
The network information is summarized in Table \ref{tab:2}, and it
includes for each network four instances of measurement placement
(denoted by A, B, C, and D) with increasing number of PMU devices;
columns 4 to 7 show the number of SCADA measurements, voltage PMU
measurements, current PMU measurements, and zero injection measurements
(when employed); the complete data sets are available for download
from \cite{CEC}. The corresponding percentage standard deviation
of the measurements are given in Table \ref{tab:1} together with
the weights. The error is simulated as Gaussian noise with zero mean
and a per-unit standard deviation computed as a percentage of the
meter full-scale reading. Note that the weights of the zero injection
measurements are needed for the CNE estimator, but not for the CEC
and REC implementations that have the zero injection measurements
handled as equality constraints. 

Table \ref{tab:3} shows the sparse matrix information and computational
performance of CEC relative to CNE for the case without zero injection
measurements, and Table \ref{tab:4} for the case with zero injection
measurements. The sparse matrix information for the CNE/CEC estimator
includes the dimension of the gain/Jacobian matrix and its number
of upper diagonal matrix non-zeros (NZ). For the case without zero
injection measurements (Table \ref{tab:3}), the size and number of
non-zeros of the CEC estimator are greater than the corresponding
CNE quantities by only 1, as the only equality constraint corresponds
to the slack angle equation. The CNE estimator computational results
in Table \ref{tab:3} correspond to the same network and measurement
sets in \cite{Dzafic_2018}, but averaged over 200 simulations of
Gaussian noise; the CEC results show a speed-up factor (SUF = time
CNE/time CEC) that approaches 1.57. For the estimation results with
zero injection measurements as shown in Table \ref{tab:4}, the SUF
factor also approaches a maximum value of around 1.49 with the execution
times averaged over 200 trials. Note that the CEC estimator computing
time on the largest instance is less than 300 ms, making it suitable
for real time applications. Tables \ref{tab:3} and \ref{tab:4} do
not include a computational performance comparison with the REC method,
as similar experiments reported in \cite{Dzafic_2018} already showed
a significant computational performance advantage in favor of the
CNE estimator. 
\begin{table}
\caption{Network Measurement Sets\label{tab:1}}
\centering{}\setlength{\tabcolsep}{3pt}{%
\begin{tabular}{ccrrrrrrr}
\toprule 
\multirow{2}{*}{Net.} &  & \multicolumn{2}{c}{Topology} &  & \multicolumn{4}{c}{Measurements}\tabularnewline
\cmidrule{3-4} \cmidrule{6-9} 
 &  & \#nodes & \#branches &  & \#SCADA & \#V-PMU & \#I-PMU & \#ZeroInj\tabularnewline
\midrule 
118\_A &  & 118 & 186 &  & 372 & 4 & 3 & 10\tabularnewline
118\_B &  & 118 & 186 &  & 372 & 3 & 35 & 10\tabularnewline
118\_C &  & 118 & 186 &  & 372 & 3 & 186 & 10\tabularnewline
118\_D &  & 118 & 186 &  & 0 & 118 & 186 & 10\tabularnewline
\midrule 
1888\_A &  & 1888 & 2531 &  & 5060 & 2 & 0 & 680\tabularnewline
1888\_B &  & 1888 & 2531 &  & 5060 & 4 & 154 & 680\tabularnewline
1888\_C &  & 1888 & 2531 &  & 5060 & 87 & 1261 & 680\tabularnewline
1888\_D &  & 1888 & 2531 &  & 0 & 1888 & 2531 & 680\tabularnewline
\midrule 
9241\_A &  & 9241 & 16049 &  & 32098 & 17 & 89 & 2901\tabularnewline
\multirow{1}{*}{9241\_B} &  & 9241 & 16049 &  & 32098 & 38 & 2007 & 2901\tabularnewline
9241\_C &  & 9241 & 16049 &  & 32098 & 66 & 8025 & 2901\tabularnewline
9241\_D &  & 9241 & 16049 &  & 0 & 9241 & 16049 & 2901\tabularnewline
\bottomrule
\end{tabular}}
\end{table}
\begin{table}
\caption{Measurement Standard Deviations and Weights \label{tab:2}}
\centering{}\setlength{\tabcolsep}{1pt}{%
\begin{tabular}{cccccccccc}
\toprule 
 & \multicolumn{3}{c}{SCADA Measurements} &  & \multicolumn{3}{c}{PMU Measurements} &  & ZI\tabularnewline
\cmidrule{2-4} \cmidrule{6-8} 
 & voltage & inj. power & power flows &  & voltage & current & ph. angle &  & Meas.\tabularnewline
\midrule
Std.Dev. & 2\% & 2\% & 2\% &  & 0.5\% & 0.5\% & $0.1^{\circ}$ &  & 0\tabularnewline
Weight & 1 & 1 & 1 &  & 5 & 5 & 5 &  & 25\tabularnewline
\bottomrule
\end{tabular}}
\end{table}
\begin{table}
\caption{Sparse Matrix information and Computational Performance of CNE/CEC
without Zero Injection Measurements (average over 200 runs) \label{tab:3}}
\centering{}\setlength{\tabcolsep}{2pt}{%
\begin{tabular}{crrrrrrrrrrrrrr}
\toprule 
\multirow{2}{*}{Net.} &  & \multicolumn{2}{c}{Matrix Size} &  & \multicolumn{2}{c}{Matrix \#NZ} &  & \multicolumn{2}{c}{\#Iteration} &  & \multicolumn{2}{c}{Time {[}ms{]}} &  & \multirow{2}{*}{SUF}\tabularnewline
\cmidrule{3-4} \cmidrule{6-7} \cmidrule{9-10} \cmidrule{12-13} 
 &  & CNE & CEC &  & CNE  & CEC &  & CNE & CEC &  & CNE & CEC &  & \tabularnewline
\midrule 
118\_A &  & 236 & 237 &  & 1052 & 1054 &  & 4.00 & 4.00 &  & 1.1 & 0.9 &  & \textbf{1.22}\tabularnewline
118\_B &  & 236 & 237 &  & 1052 & 1054 &  & 4.00 & 4.00 &  & 1.3 & 1.13 &  & \textbf{1.15}\tabularnewline
118\_C &  & 236 & 237 &  & 1052 & 1054 &  & 4.00 & 4.00 &  & 1.4 & 1.17 &  & \textbf{1.20}\tabularnewline
118\_D &  & 236 & 237 &  & 595 & 596 &  & 1.00 & 1.00 &  & 0.4 & 0.4 &  & \textbf{1.00}\tabularnewline
\midrule 
1888\_A &  & 3776 & 3777 &  & 14121 & 14123 &  & 5.11 & 5.11 &  & 26.4 & 23.4 &  & \textbf{1.13}\tabularnewline
1888\_B &  & 3776 & 3777 &  & 14121 & 14123 &  & 5.00 & 5.00 &  & 28.4 & 24.5 &  & \textbf{1.16}\tabularnewline
1888\_C &  & 3776 & 3777 &  & 14121 & 14123 &  & 4.00 & 4.00 &  & 22.3 & 18.1 &  & \textbf{1.23}\tabularnewline
1888\_D &  & 3776 & 3777 &  & 8393 & 8394 &  & 1.00 & 1.00 &  & 5.3 & 5.0 &  & \textbf{1.05}\tabularnewline
\midrule 
9241\_A &  & 18482 & 18483 &  & 82127 & 82128 &  & 5.00 & 5.00 &  & 312.7 & 189.3 &  & \textbf{1.65}\tabularnewline
9241\_B &  & 18482 & 18483 &  & 82127 & 82128 &  & 5.00 & 5.00 &  & 308.6 & 188.7 &  & \textbf{1.64}\tabularnewline
9241\_C &  & 18482 & 18483 &  & 82127 & 82128 &  & 5.00 & 5.00 &  & 311.4 & 198.8 &  & \textbf{1.57}\tabularnewline
9241\_D &  & 18482 & 18483 &  & 46897 & 46898 &  & 1.00 & 1.00 &  & 50.3 & 45.2 &  & \textbf{1.11}\tabularnewline
\bottomrule
\end{tabular}}
\end{table}
\begin{table}
\caption{Sparse Matrix information and Computational Performance of CNE/CEC
with Zero Injection Measurements (average over 200 runs) \label{tab:4}}
\centering{}\setlength{\tabcolsep}{1.5pt}{%
\begin{tabular}{lrrrrrrrrrrrrrr}
\toprule 
\multirow{2}{*}{Net.} &  & \multicolumn{2}{c}{Matrix Size} &  & \multicolumn{2}{c}{Matrix \#NZ} &  & \multicolumn{2}{c}{\#Iteration} &  & \multicolumn{2}{c}{Time {[}ms{]}} &  & \multirow{2}{*}{SUF}\tabularnewline
\cmidrule{3-4} \cmidrule{6-7} \cmidrule{9-10} \cmidrule{12-13} 
 &  & CNE & CEC &  & CNE  & CEC &  & CNE & CEC &  & CNE & CEC &  & \tabularnewline
\midrule 
118\_A &  & 236 & 257 &  & 1146 & 1162 &  & 4.00 & 4.00 &  & 1.46 & 1.42 &  & \textbf{\textit{1.03}}\tabularnewline
118\_B &  & 236 & 257 &  & 1146 & 1162 &  & 4.00 & 4.00 &  & 1.50 & 1.46 &  & \textbf{\textit{1.03}}\tabularnewline
118\_C &  & 236 & 257 &  & 1146 & 1162 &  & 3.96 & 3.95 &  & 1.65 & 1.54 &  & \textbf{\textit{1.07}}\tabularnewline
118\_D &  & 236 & 257 &  & 759 & 704 &  & 3.00 & 3.00 &  & 0.88 & 0.84 &  & \textbf{\textit{1.05}}\tabularnewline
\midrule 
1888\_A &  & 3776 & 5137 &  & 20960 & 21106 &  & 5.00 & 5.00 &  & 49.17 & 45.53 &  & \textbf{\textit{1.08}}\tabularnewline
1888\_B &  & 3776 & 5137 &  & 20960 & 21106 &  & 5.00 & 5.00 &  & 5.48 & 45.96 &  & \textbf{\textit{1.12}}\tabularnewline
1888\_C &  & 3776 & 5137 &  & 20960 & 21081 &  & 5.00 & 5.00 &  & 55.40 & 47.76 &  & \textbf{\textit{1.16}}\tabularnewline
1888\_D &  & 3776 & 5137 &  & 18971 & 15378 &  & 4.00 & 4.00 &  & 33.93 & 31.71 &  & \textbf{\textit{1.07}}\tabularnewline
\midrule 
9241\_A &  & 18482 & 24285 &  & 100903 & 108594 &  & 5.00 & 5.00 &  & 381.85 & 267.30 &  & \textbf{\textit{1.43}}\tabularnewline
9241\_B &  & 18482 & 24285 &  & 100903 & 108654 &  & 5.00 & 5.00 &  & 405.85 & 272.54 &  & \textbf{\textit{1.49}}\tabularnewline
9241\_C &  & 18482 & 24285 &  & 100903 & 108716 &  & 5.00 & 5.00 &  & 412.61 & 283.67 &  & \textbf{\textit{1.45}}\tabularnewline
9241\_D &  & 18482 & 24285 &  & 80116 & 73646 &  & 4.00 & 4.00 &  & 247.12 & 180.74 &  & \textbf{\textit{1.37}}\tabularnewline
\bottomrule
\end{tabular}}
\end{table}
\begin{table}
\caption{Measurement Performance Accuracy Indices with Zero Injection Measurements
(average over 200 runs)\label{tab:5}}
\centering{}\setlength{\tabcolsep}{6pt}{%
\begin{tabular}{lrrrrr}
\toprule 
\multirow{1}{*}{Net.} & CEC  & CNE & PIF-CNE & REC & PIF-REC\tabularnewline
\midrule 
118\_C & 0.043730 & 0.043903 & \textbf{\textit{1.004}} & 0.048352  & \textbf{\textit{1.106}}\tabularnewline
118\_D & 0.006822 & 0.007237 & \textbf{\textit{1.061}} & 0.008464 & \textbf{\textit{1.241}}\tabularnewline
\midrule 
1888\_C & 0.180213 & 0.184422 & \textbf{\textit{1.034}} & 0.263175 & \textbf{\textit{1.460}}\tabularnewline
1888\_D & 0.002982 & 0.006496 & \textbf{\textit{2.178}} & 0.008288 & \textbf{\textit{2.779}}\tabularnewline
\midrule 
9241\_C & 0.133647 & 0.154353 & \textbf{\textit{1.155}} & 0.182634 & \textbf{\textit{1.367}}\tabularnewline
9241\_D & 0.001161 & 0.003526 & \textbf{\textit{3.037}} & 0.004220 & \textbf{\textit{3.635}}\tabularnewline
\bottomrule
\end{tabular}}
\end{table}

In addition to the AVX-2 CEC implementation being faster than the
recent CNE implementation in \cite{Dzafic_2018}, the CEC estimator
is also more accurate than the CNE due to the exact modeling of zero
injection measurements. \textcolor{black}{The accuracy is quantified
using performance indices for the measurement error (\ref{eq:performance_index_2})
and the voltage error (\ref{eq:performance_index_1}) \cite{Valverde_2011}:}

{\small{}
\begin{equation}
\xi_{z}=\frac{\sum_{i=1}^{m}\left|z_{i}^{estimated}-z_{i}^{true}\right|^{2}}{\sum_{i=1}^{m}\left|z_{i}^{measured}-z_{i}^{true}\right|^{2}}\label{eq:performance_index_2}
\end{equation}
}{\small \par}

{\small{}
\begin{equation}
\sigma_{x}^{2}=\sum_{i=\text{1}}^{n}\left|x_{i}^{estimated}-x_{i}^{true}\right|^{2}\label{eq:performance_index_1}
\end{equation}
}{\small \par}

For estimation with zero injection measurements, Tables \ref{tab:5}
and \ref{tab:6} respectively show the performance indices for the
measurement error and voltage error. For each network in these tables,
the performance indices are computed after the state vector is estimated
via the CEC, the CNE \cite{Dzafic_2018}, and the REC \cite{Valverde_2011}
methods. Two performance improvement factor ({\small{}$\mathrm{PIF}$})
ratios are used to quantify how the measurement (\ref{eq:performance_index_2})
and voltage (\ref{eq:performance_index_1}) performance indices of
the CNE and REC estimators compare against CEC; these ratios are: 

{\small{}
\begin{equation}
\mathrm{PIF{-}CNE=\frac{performance\:index\:of\:CNE}{performance\:index\:of\:CEC}}
\end{equation}
}{\small \par}

{\small{}
\begin{equation}
\mathrm{PIF{-}REC=\frac{performance\:index\:of\:REC}{performance\:index\:of\:CEC}}
\end{equation}
}{\small \par}

The results in Tables \ref{tab:5} and \ref{tab:6} show that both
the {\small{}$\mathrm{PIF-CNE}$} and {\small{}$\mathrm{PIF-REC}$}
indices are consistently greater than 1, and that the performance
improvement can be excess of 3 when evaluated for measurement accuracy.
The CEC estimator also exhibits superior computational performance
under stressed conditions leading to voltage instability \cite{Venikov_1975},
as evidenced by the results in Table \ref{tab:7}; in this table,
the load of the 1118\_A test instance is uniformly increased and both
the CEC and REC methods are used to estimate the state. At the highest
load multiplication factor of 1.077, the CEC estimator required 7
iterations to converge while the REC required 10; the corresponding
convergence pattern is shown in Table \ref{tab:8}. The CNE estimator
therefore improves precision with reduced computational requirements;
this makes it suitable in applications for improving the quality of
state estimation \cite{Angelos_2016} and enhancing the processing
of bad data \cite{BrownFilho_2014}.
\begin{table}
\caption{Voltage Performance Accuracy Indices with Zero Injection Measurements
(average over 200 runs)\label{tab:6}}
\centering{}\setlength{\tabcolsep}{6pt}{%
\begin{tabular}{lrrrrr}
\toprule 
\multirow{1}{*}{Net.} & CEC  & CNE & PIF-CNE & REC & PIF-REC\tabularnewline
\midrule 
118\_C & 0.000936 & 0.000951 & \textbf{\textit{1.016}} & 0.001225 & \textbf{\textit{1.309}}\tabularnewline
118\_D & 0.000074 & 0.000079 & \textbf{\textit{1.068}} & 0.000114  & \textbf{\textit{1.541}}\tabularnewline
\midrule 
1888\_C & 0.032794 & 0.033116 & \textbf{\textit{1.010}} & 0.053508 & \textbf{\textit{1.632}}\tabularnewline
1888\_D & 0.004831 & 0.005756 & \textbf{\textit{1.191}} & 0.008754 & \textbf{\textit{1.812}}\tabularnewline
\midrule 
9241\_C & 0.092998  & 0.099949 & \textbf{\textit{1.075}} & 0.118473 & \textbf{\textit{1.274}}\tabularnewline
9241\_D & 0.020224 & 0.022186 & \textbf{\textit{1.097}} & 0.023807 & \textbf{\textit{1.177}}\tabularnewline
\bottomrule
\end{tabular}}
\end{table}
\begin{table}
\caption{Numerical Stability Test of the CEC and REC Estimators on the 1888\_A
Network Instance ($\varepsilon=10^{-7}$) \label{tab:7}}
\centering{}\setlength{\tabcolsep}{8pt}{%
\begin{tabular}{rrrrrrr}
\toprule 
\multirow{2}{*}{Mult.} &  & \multicolumn{2}{r}{Min. Voltage} &  & \multicolumn{2}{c}{\#Iter.}\tabularnewline
\cmidrule{3-4} \cmidrule{6-7} 
 &  & Node No. & Value &  & CEC  & REC\tabularnewline
\midrule 
1.000 &  & 1194 & 0.83151 &  & 5 & 5\tabularnewline
1.050 &  & 1194 & 0.81383 &  & 6 & 6\tabularnewline
1.054 &  & 1194 & 0.81177 &  & 7 & 8\tabularnewline
1.070 &  & 1194 & 0.80110 &  & 7 & 8\tabularnewline
1.072 &  & 1194 & 0.79914 &  & 7 & 9\tabularnewline
1.077 &  & 358 & 0.76614 &  & 7 & 10\tabularnewline
\bottomrule
\end{tabular}}
\end{table}
\begin{table}
\caption{Convergence Pattern of the CEC and REC Estimators on the 1888\_A Network
Instance with a Load Multiplier of 1.077 ($\varepsilon=10^{-7}$)\label{tab:8}}
\centering{}\setlength{\tabcolsep}{8pt}{%
\begin{tabular}{crr}
\toprule 
\multirow{2}{*}{Iteration} & \multicolumn{2}{c}{Precision}\tabularnewline
\cmidrule{2-3} 
 & CEC  & REC\tabularnewline
\midrule 
1 & 1.48116e+00 & 1.48116e+00\tabularnewline
2 & 1.19273e+00 & 4.22265e+00\tabularnewline
3 & 5.41629e-01 & 6.03061e-01\tabularnewline
4 & 1.71176e-01 & 1.37050e+00\tabularnewline
5 & 1.01962e-02 & 1.61155e+00\tabularnewline
6 & 1.45915e-04 & 3.87409e-01\tabularnewline
7 & 1.03513e-08 & 5.85039e-02\tabularnewline
8 &  & 1.33321e-03\tabularnewline
9 &  & 2.62295e-07\tabularnewline
10 &  & 3.20996e-13\tabularnewline
\bottomrule
\end{tabular}}
\end{table}

\section{Conclusion \label{sec:Conclusion}}

This paper presented an algorithm for direct hybrid state estimation
using a complex equality constrained normal equations approach. The
complex variable formulation is advantageous for handling PMU measurements,
and it is naturally suited for implementation on modern processors
that allow fused multiply-accumulate operations and closely related
advances. The use of equality constraints permit accurate modeling
of zero injection measurements, and it has demonstrable benefits on
the state estimator performance indices. The implementation of the
CEC estimator is reported using the advanced vector extensions (AVX-2)
set of instructions, which allows faster and specialized operations
on complex numbers. Numerical results are reported on large scale
transmission networks having different SCADA and PMU measurement configurations,
and the results indicate that the AVX-2 implementation on networks
with 9241 nodes requires less than 300 ms, thus conforming with real-time
computing requirements. A comparison is carried out with two hybrid
state  estimation techniques: the complex normal equations (CNE) approach
and the real equality constrained (REC) estimator; the comparative
analysis shows that the proposed AVX-2 CEC implementation is superior
to both methods in terms of solution speed and accuracy. \balance

\appendix{}

\subsubsection*{Complex Zero Injection Measurements}

Consider the complex zero injection power at node {\small{}$i$} and
its conjugate: 

{\small{}
\begin{equation}
\begin{aligned}s_{i} & =0=u_{i}\overline{Y}_{ii}\overline{u}_{i}+u_{i}\stackrel[\underset{k\neq i}{k=1}]{n}{\sum}\overline{Y}_{ik}\overline{u}_{k}\\
\overline{s}_{i} & =0=\overline{u}_{i}Y_{ii}u_{i}+\overline{u}_{i}\stackrel[\underset{k\neq i}{k=1}]{n}{\sum}Y_{ik}u_{k}
\end{aligned}
\end{equation}
}where{\small{}
\begin{equation}
Y_{ii}=y_{i}^{sh}+\sum_{\underset{k\neq i}{k=1}}^{n}y_{ik},\quad Y_{ik}=-y_{ik},\quad k\neq i
\end{equation}
$y_{ik}$} is the series admittance of branch {\small{}$ik$}, {\small{}$y_{i}^{sh}$}
is the shunt admittance at node {\small{}$i$}, {\small{}$u_{i}$}
is the phasor voltage at node {\small{}$i$}, and {\small{}$n$} is
the number of nodes. The corresponding elements of the Jacobian and
conjugate Jacobian elements in {\small{}$J$} are:

{\small{}
\begin{equation}
\begin{aligned}\frac{\partial s_{i}}{\partial u_{i}} & =\overline{Y}_{ii}\overline{u}_{i}+\stackrel[\underset{k\neq i}{k=1}]{n}{\sum}\overline{Y}_{ik}\overline{u}_{k}\\
\frac{\partial s_{i}}{\partial u_{k}} & =0,\quad k\neq i\\
\frac{\partial s_{i}}{\partial\overline{u}_{i}} & =\overline{Y}_{ii}u_{i}\\
\frac{\partial s_{i}}{\partial\overline{u}_{k}} & =\overline{Y}_{ik}u_{i},\quad k\neq i
\end{aligned}
\end{equation}
\begin{equation}
\begin{aligned}\frac{\partial\overline{s}_{i}}{\partial u_{i}} & =Y_{ii}\overline{u}_{i}=\overline{\left(\frac{\partial s_{i}}{\partial\overline{u}_{i}}\right)}\\
\frac{\partial\overline{s}_{i}}{\partial u_{k}} & =Y_{ik}\overline{u}_{i}=\overline{\left(\frac{\partial s_{i}}{\partial\overline{u}_{k}}\right)},\quad k\neq i\\
\frac{\partial\overline{s}_{i}}{\partial\overline{u}_{i}} & =Y_{ii}u_{i}+\stackrel[\underset{k\neq i}{k=1}]{n}{\sum}Y_{ik}u_{k}=\overline{\left(\frac{\partial s_{i}}{\partial u_{i}}\right)}\\
\frac{\partial\overline{s}{}_{i}}{\partial\overline{u}_{k}} & =0,\quad k\neq i
\end{aligned}
\end{equation}
}{\small \par}

\subsubsection*{Slack Angle}

The slack angle condition requires setting the imaginary part of the
slack node voltage to zero:

{\small{}
\begin{equation}
u_{s}^{im}=\frac{\mathtt{i}}{2}\left(\overline{u}_{s}-u_{s}\right)=0
\end{equation}
}The corresponding Jacobian and conjugate Jacobian elements in {\small{}$J$}
are:

{\small{}
\begin{align}
\frac{\partial u_{s}^{im}}{\partial u_{s}} & =-\frac{\mathtt{i}}{2}\\
\frac{\partial u_{s}^{im}}{\partial\overline{u}_{s}} & =\frac{\mathtt{i}}{2}=\overline{\left(\frac{\partial u_{s}^{im}}{\partial u_{s}}\right)}
\end{align}
}{\small \par}

\bibliographystyle{IEEEtran}
\bibliography{Constr_CSE_refs}

\vspace{0\baselineskip} 
\begin{IEEEbiographynophoto}{Izudin Džafi\'{c}}
 \textcolor{black}{(M'05-SM'13) received his Ph.D. degree from University
of Zagreb, Croatia in 2002. He is currently a Professor in the Department
of Electrical Engineering at the International University of Sarajevo,
Bosnia. From 2002 to 2014, he was with Siemens AG, Nuremberg, Germany,
where he held the position of the Head of the Department and Chief
Product Owner (CPO) for Distribution Network Analysis (DNA) R\&D.
His research interests include power system modeling, development
and application of fast computing to power systems simulations. Dr.
Džafi\'{c} is a member of the IEEE Power and Energy Society and the
IEEE Computer Society.}
\end{IEEEbiographynophoto}

\vspace{0\baselineskip} 
\begin{IEEEbiographynophoto}{Rabih Jabr}
 (M'02-SM'09-F'16) was born in Lebanon. He received the B.E. degree
in electrical engineering (with high distinction) from the American
University of Beirut, Beirut, Lebanon, in 1997 and the Ph.D. degree
in electrical engineering from Imperial College London, London, U.K.,
in 2000. Currently, he is a Professor in the Department of Electrical
and Computer Engineering at the American University of Beirut. His
research interests are in mathematical optimization techniques and
power system analysis and computing. 
\end{IEEEbiographynophoto}

\vspace{0\baselineskip} 
\begin{IEEEbiographynophoto}{Bikash C. Pal}
 (M'00-SM'02-F'13) received the B.E.E. (with honors) degree from
Jadavpur University, Calcutta, India, the M.E. degree from the Indian
Institute of Science, Bangalore, India, and the Ph.D. degree from
Imperial College London, London, U.K., in 1990, 1992, and 1999, respectively,
all in electrical engineering. Currently, he is a Professor in the
Department of Electrical and Electronic Engineering, Imperial College
London. His current research interests include state estimation, power
system dynamics, and flexible ac transmission system controllers.
\end{IEEEbiographynophoto}

\end{document}